\documentclass[aps,pre,showpacs]{revtex4}
\usepackage{graphicx}
\usepackage{epsfig}
\usepackage{amsmath}
\usepackage{graphics}
\usepackage{latexsym}
\usepackage{subfigure}
\usepackage{color}

\begin{document}

\title{Do the contact angle and line tension of surface-attached droplets depend on the radius of curvature?
}

\author{Subir K. Das $^1$, Sergei A. Egorov $^{2,3}$, Peter Virnau $^4$, David Winter $^4$, and Kurt Binder $^4$}
\affiliation{$^1$ Theoretical Sciences Unit, Jawaharlal Nehru Centre for Advanced Scientific Research, Jakkur, Bangalore, 56004 India}

\affiliation{$^2$ Department of Chemistry, University of Virginia, Charlottesville, USA}

\affiliation{$^3$ Leibniz Institut f\"{u}r Polymerforschung Dresden, Hohe Strasse 6, D-01069 Dresden, Germany}

\affiliation{$^4$ Institut f\"ur Physik, Johannes Gutenberg-Universit\"at Mainz, Staudinger Weg 9, 55099 Mainz, Germany}





\begin{abstract}
Results from Monte Carlo simulations of wall-attached droplets in the three-dimensional Ising lattice gas
model and in a symmetric binary Lennard-Jones fluid, confined by antisymmetric walls,
are analyzed, with the aim to estimate the dependence of the contact angle $(\Theta)$
on the droplet radius $(R)$ of curvature. Sphere-cap shape of the wall-attached droplets
is assumed throughout. An approach,  based purely on ``thermodynamic'' observables,
e.g., chemical potential, excess density due to the droplet, etc., is used, to avoid ambiguities in
the decision which particles belong (or do not belong, respectively) to the droplet. It
is found that the results are compatible with a variation $[\Theta(R)-\Theta_{\infty}] \propto 1/R$,
$\Theta_{\infty}$ being the contact angle in the thermodynamic limit ($R=\infty$).
The possibility to use such results to estimate the excess free energy related to the contact line of
the droplet, namely the line tension, at the wall, is discussed.
Various problems that hamper this approach and were
not fully recognized in previous attempts to extract the line tension are identified. It is
also found that the dependence of wall tensions on the difference of chemical potential of the droplet
from that at the bulk coexistence provides effectively a change of the contact angle of similar magnitude.
The simulation approach yields precise estimates for the excess density due to wall-attached droplets
and the corresponding free energy excess, relative to a system without a droplet at the same chemical
potential. It is shown that this information suffices to estimate nucleation barriers, not affected by
ambiguities on droplet shape, contact angle and line tension.

 \end{abstract}

\maketitle
e-mail address: das@jncasr.ac.in
%
%
%
%
%

\section{Introduction}

Fluid nano-droplets on planar substrates are important in the context of
heterogeneous nucleation \cite{1,2,3,4,28} and of much recent interest \cite{nr1,nr2,nr3},
given that these have diverse applications in connection
with microfluidics, nanofluidics, lithography, etc. \cite{5,6}. Their properties
in (metastable) equilibrium are controlled by a subtle interplay of various
surface tensions and the excess free energy attributed to the three phase contact
line, the so-called line tension ($\tau$)
\cite{7,8,9,10,11,12,13,14,15} (see Fig.~\ref{fig1} for schematic depiction).
However, theoretical and conceptual aspects of line tension are still controversially
discussed (e.g. \cite{10,11,12,13}) and attempts to estimate it
experimentally have often led to results whose validity is still
debated \cite{9}.

In this work we shall study various aspects related to wall-attached droplets
via computer simulations, as previously done, see e.g. Refs. \cite{16,17,18,19,20,21,22,23,49}. Unlike
our previous works \cite{18,19,20,49}, we pay attention to the fact that for
nano-droplets, in heterogeneous context, the contact angle ($\Theta$) may depend on the droplet radius,
e.g. because of the line tension \cite{15,24,25}. This dependence needs to be
taken into account self-consistently when one tries to estimate the line tension.
Even if the line tension is disregarded, the contact angle of a nano-droplet may
differ from its macroscopic value ($\Theta_{\infty}$) for other reasons \cite{11}. This problem can
be avoided, at least in the framework of models possessing the
particle-hole symmetry, like the Ising lattice gas system,
between bulk coexisting phases, by considering a simulation geometry
(shown in Fig.~\ref{fig2}) \cite{23} involving inclined planar interfaces between
``antisymmetric'' walls. We have demonstrated the viability of this approach for
estimating contact angles also by lattice-based self-consistent field theory (SCFT),
see e.g. \cite{26}, for a symmetric binary mixture between two antisymmetric walls.
However, as stated above, the objective of the present work is to explore the curvature
dependence.

The sketch drawn in Fig.~\ref{fig1}(a) implicitly uses a description appropriate for
macroscopic droplets, such as water droplets on a car window visible by the naked eye:
the droplet surface is drawn infinitely thin, and the contact line where it hits the wall
is well-defined. However, it is doubtful that such a description can still be applied at the
nanoscale without ambiguities resulting from the fact that on molecular scales interfaces
are diffuse, there is no well-defined contact line then, and the nano-droplet shape may differ
from the sphere-cap shape appropriate in the macroscopic limit (in the absence of gravity). In
this work, we shall present a simulation approach that provides the possibility to extract
properties such as contact angles, line tensions, etc., for liquid nanodroplets. At the same time,
we shall demonstrate that it is possible to obtain estimates for the free energy barrier for
the (heterogeneous) nucleation of such droplets and these estimates do not
suffer from the ambiguities resulting from situation
when the macroscopic description invoked in Fig.~\ref{fig1} (a)
is taken too literally.

In Sec. II we shall concisely summarize the theoretical background of our work.
In Sec. III we briefly describe our approach for the Ising lattice gas model, before showing that the
resulting estimates for the contact angles indeed differ significantly from previous
work \cite{18,19}, exhibiting a reasonably pronounced dependence on the droplet radius $R$. This
dependence is qualitatively consistent with standard predictions \cite{15,24} 
of a correction of order $\tau/R$, when line tension
estimates for planar interfaces \cite{23} (see Fig.~\ref{fig2}) are inserted. Due to
anisotropy effects from the lattice, quantitative conclusions from our work are delicate, however.
In addition, the lack in accuracy of the data \cite{18,19} does not allow us to obtain
significantly improved estimates for the line tension.  Sec. IV presents corresponding
results for a symmetric binary ($A,B$) Lennard-Jones mixture, along with a description of the model and methods.
Again quite noticeable differences with respect to the approximate approach of the previous work
\cite{20} are found. In this case also the results can be interpreted as an effective
radius dependence of the contact angle, as mentioned above. Sec. V then discusses estimates for
the $\delta \mu$-dependence of contact angle, obtained via a mean-field type
approach \cite{26}, coming via the wall tensions,
for a lattice model of a binary mixture. Finally we
summarize our findings in Sec. VI.

\section{THEORETICAL BACKGROUND}

\begin{figure}[ht]
\subfigure[]{
\includegraphics[width=8cm,clip]{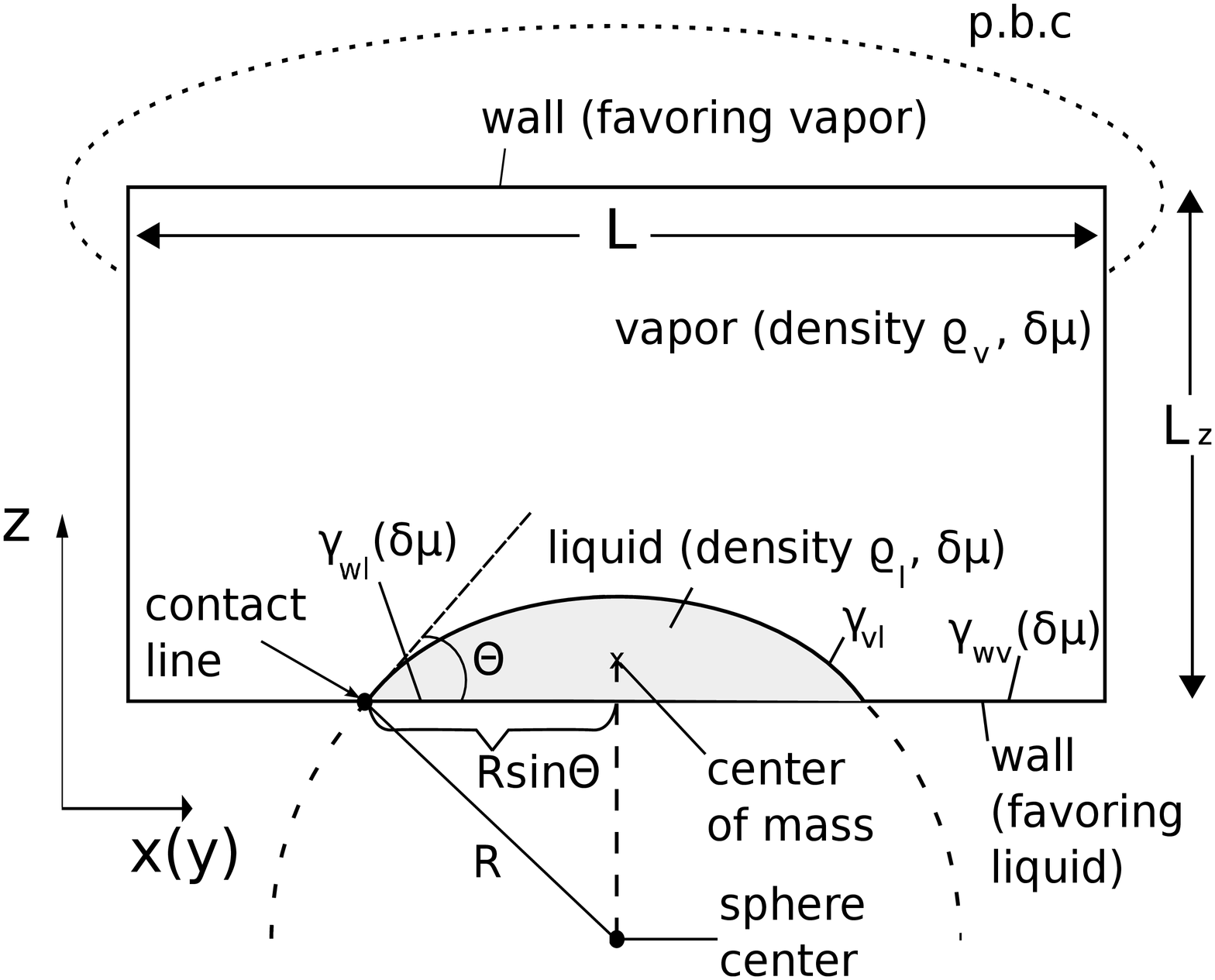}
}
\subfigure[]{
\includegraphics[width=8cm,clip]{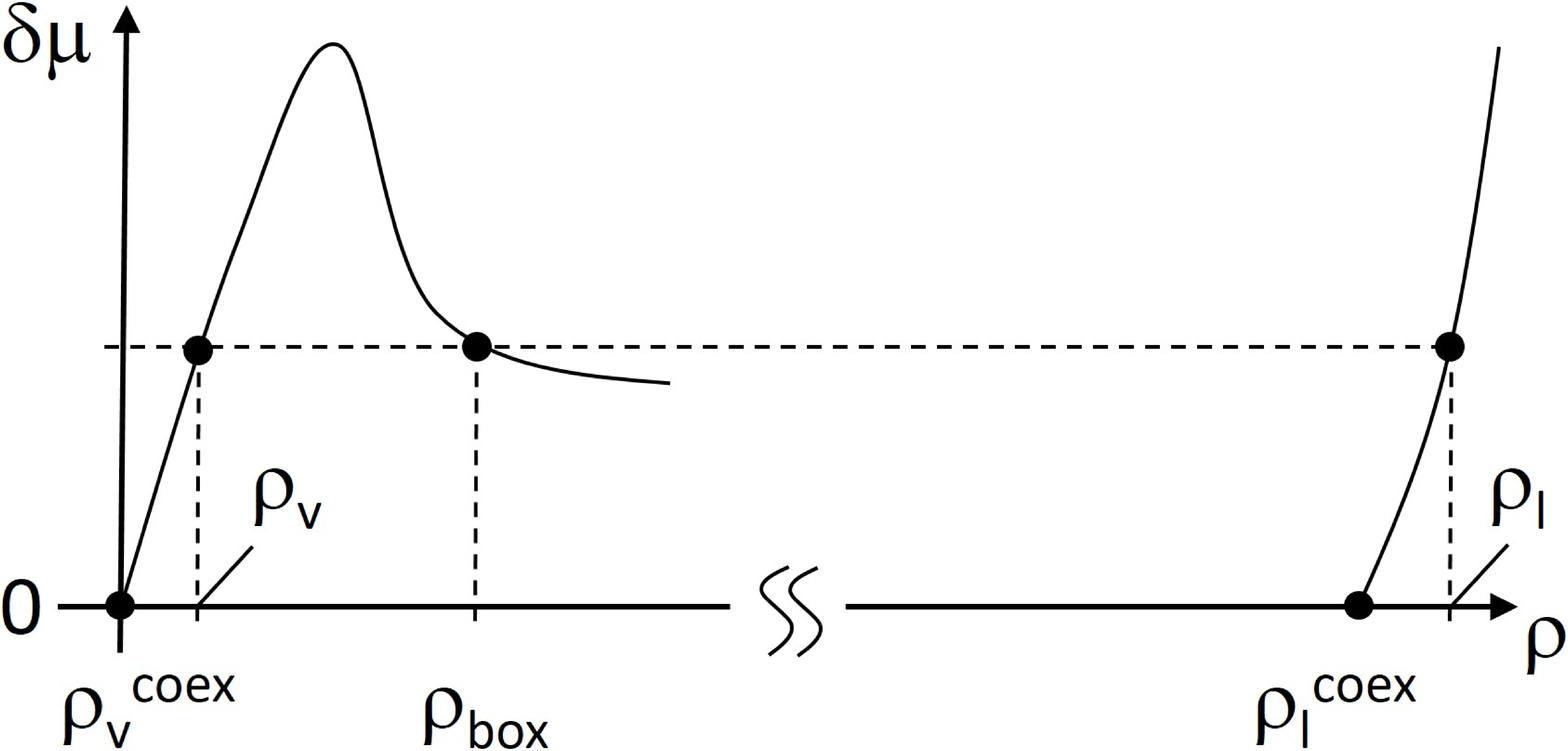}
}
\centering
\caption{\label{fig1} (a) Simulation geometry used for the study of wall-attached
sphere-cap shaped droplets in the Ising lattice gas model, showing schematically
a cross section through the droplet center of mass (marked by a cross) in the
$zx$-plane. The lattice has linear dimensions $L$ in $x$ and $y$ direction and
$L_z$ in $z$-direction, with periodic boundary conditions (p.b.c.) in $x$ and $y$
directions only. At the lower wall [first plane ($n=1$) of the simple cubic lattice
in $z$-direction] a positive surface field $H_1$ acts and at the upper wall
($n=L_z)$ a negative surface field $H_{L_z}=-H_1$ acts, so that an antisymmetric boundary
condition is created. Otherwise, free boundary conditions in $z$-direction are used
(i.e, missing spins in planes $n=0$ and $n=L_z+1$). The radius of curvature $R$
and the contact angle $\Theta$ of the droplet are indicated. The contact line
meeting with the shown plane is indicated by a full dot (only on the left side). Note that the shown
state is in stable equilibrium for a chemical potential $\mu=\mu_{\rm coex} +\delta \mu$,
where $\mu_{\rm coex}$ is the chemical potential at bulk coexistence, with vapor
and liquid densities $\rho_v^{\rm coex}$ and $\rho^{\rm coex}_\ell$,
respectively. (b) Chemical potential
difference $\delta\mu$ plotted vs. density $\rho$, for the geometry of part (a) which
refers to a choice of the density $\rho=\rho_{\rm box}$ to the right of the peak
(related to the droplet evaporation/condensation transition, and is
not of interest here). For finite size of the radius $R$ in (a), $\delta \mu > 0$
in equilibrium, and correspondingly $\rho_v$ and $\rho_{\ell}$ are enhanced in comparison
with their corresponding values at two-phase coexistence in the bulk.} \end{figure}

\begin{figure} [ht]
\includegraphics[width=8cm,clip]{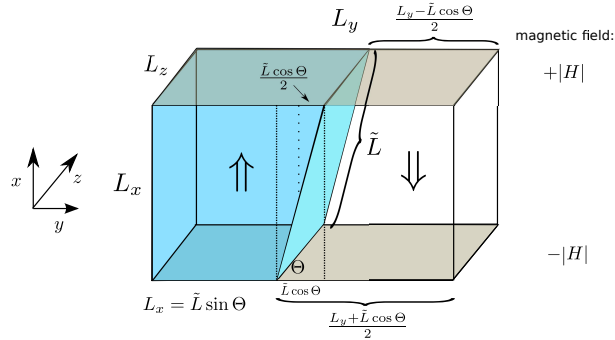}
\caption{\label{fig2} Sketch of the geometry used to obtain the ``macroscopic'' line tension
for the Ising model, considering a simple cubic lattice with linear dimensions $L_x,L_y$ and
$L_z$ in the $x,y$, $z$ directions, respectively, in the limit where all these linear dimensions
become macroscopically large. While a periodic boundary condition is used in the
$z$-direction, in the $x$-direction two free surfaces of linear dimensions
$L_y \times L_z$ are used, at which boundary fields $H_1=-|H_1|$ at the bottom
and $H_n=+|H_1|$ at the top act. When one uses a generalized anti-periodic boundary
condition (GAPBC), see \cite{23}, phase coexistence between two domains with opposite
magnetization $+m_{\rm coex}$ and $-m_{\rm coex}$ in the bulk (symbolized by the two double arrows)
occurs. These domains are separated by a planar interface that is inclined by an angle
$\Theta$ with respect to the $yz$-plane.} \end{figure}

 We now discuss the situation sketched in Fig.~\ref{fig1}(a) in more detail.
Note that in the thermodynamic limit, when the droplet radius of curvature
$R \rightarrow \infty$, stable phase-coexistence between a liquid droplet
(at density $\rho^{\rm coex}_\ell)$ and surrounding vapor (at density
$\rho^{\rm coex}_v)$ is possible in the bulk. Then, of course, the chemical potential
difference ($\delta\mu = \mu - \mu_{\rm coex}$) between the (liquid) droplet and vapor (at bulk coexistence)
is identically zero. As opposed to the spherical shape in the bulk, the droplet has
sphere-cap shape at the surface (or wall) and the related contact angle
is given by Young's equation \cite{8,27}
\begin{equation} \label{eq1}
\gamma_{\ell v} \cos \Theta_\infty=\gamma_{wv} (\delta \mu) -\gamma_{w \ell} (\delta \mu),
\quad \delta \mu=0.
\end{equation}
Here  $\gamma_{\ell v}$ is the tension related to a planar liquid-vapor interface,
while $\gamma_{w v} (\delta \mu)$ and $\gamma_{w \ell} (\delta \mu)$ are the surface
tensions of vapor and liquid against the wall \cite{23}.

In the situation  of interest, however, where $R$ is finite, we need
to distinguish $\Theta$ from $\Theta_\infty$. 
If we consider the droplet of Fig.~\ref{fig1} (a) to be nanoscopic, the chemical potential
difference $\delta \mu$ is positive [see the qualitative demonstration in Fig. \ref{fig1} (b)].
The wall tensions $\gamma_{w v} (\delta \mu)$ and
$\gamma_{ w \ell} (\delta \mu)$, in that case, will, in general, differ from their counterparts for
$\delta \mu=0$ -- see Ref. \cite{11}. In the framework of Monte Carlo (MC) simulations, it is easily
possible to compute these wall tensions with the variation of $\delta \mu$ (or of densities $\rho_v$
and $\rho_\ell$, respectively) \cite{29,30}. At least for the model studied in
Ref. \cite{29,30} the dependence of $\gamma_{w v}$ and $\gamma_{w \ell}$ on density is
rather pronounced. Since $\gamma_{w v}$ decreases with density, while
$\gamma_{w \ell}$ increases, this effect does not cancel out in the difference
$\gamma_{w v} - \gamma_{ w \ell}$, which matters in Young's equation. In addition,
the excess free energy due to the
line tension \cite{7,8,9,10,11,12,13,14,15} plays crucial role. Some authors
\cite{24,31}, hence, argued that Eq.~(\ref{eq1}) should be replaced by
\begin{equation} \label{eq2}
\gamma_{\ell v} \cos  \Theta=\gamma_{wv} (\delta \mu) -
\gamma_{w \ell} (\delta \mu) - \frac{\tau}{R \sin \Theta}.
\end{equation}

Eq.~(\ref{eq2}) seems to be almost obvious as a consequence of the mechanical force
balance at the contact line \cite{15}. However, the real situation is not as simple, given that
all interfaces are diffuse objects. In that situation, the notion of a contact line (particularly
its length) may be influenced by the use of somewhat arbitrarily chosen ``dividing surfaces''
for the interfaces \cite{10}. Even for a spherical droplet in the bulk, the ``equimolar
dividing surface'' needs to be distinguished from the ``surface of tension'' \cite{8}.
Naturally, the length of the contact line, for a sphere cap, will depend on the choice of a dividing
surface \cite{10,11}.
In any case, when the dependence of the wall tensions on $\delta \mu$ can be neglected,
Eq.~(\ref{eq2}) can be reduced to the simpler result \cite{15}
\begin{equation} \label{eq3}
\frac{\tau}{R} = \gamma_{\ell v} \sin \Theta (\cos \Theta_\infty - \cos \Theta).
\end{equation}

Now we remind the reader of the basic
concepts of the theory of heterogeneous nucleation \cite{1,2,3,4}. There one does not consider a small
finite size system with linear dimensions comparable to the droplet radius as drawn in Fig.~\ref{fig1}.
Systems of interest, in fact, are
macroscopic ones, being in a metastable vapor phase $(\delta \mu > 0$ so that the liquid
would be the stable phase), 
exposed to a wall in the grand-canonical ensemble. Then, $\delta \mu$, rather than the
density $\rho$ in the system, is chosen as an independent variable. We emphasize at the outset that this
will not be done in the simulations of the later sections. We will use the density, 
rather than the chemical potential,
as the given independent variable.

Then one asks the question, assuming that a wall-attached sphere-cap shaped droplet with radius of curvature $R$ and contact
angle $\Theta$ forms, what would be its Gibbs free energy cost $\Delta F(R)$. Note that this question refers to a somewhat
hypothetical situation, since such droplets are intrinsically unstable, out of equilibrium, and hence the use of equilibrium-type
descriptions is somewhat questionable. However, in the end one
is mostly interested in the position $R^*$ of the free energy maximum and its value $\Delta F (R^*)$ only.

Hence, the Gibbs excess free energy of the droplet in Fig.~\ref{fig1} (a),
relative to a state of the system at density $\rho_v$ [see Fig. \ref{fig1} (b)], without a droplet, becomes
(for small $\delta \mu$)
\begin{equation} \label{eq6}
\Delta F(R)=-V_d \delta \mu (\rho_\ell ^{\rm coex} - \rho^{\rm coex}_v) + A \gamma_{\ell v} + A_b ( \gamma_{w \ell} (\delta \mu) - \gamma_{wv} (\delta \mu)) + \ell \tau.
\end{equation}
where $\ell=2 \pi R \sin \Theta$ is the length of the contact line.
Furthermore, simple geometric considerations yield the volume $V_d$, upper surface area $A$ and
basal area $A_b$ of the droplet as
\begin{equation} \label{eq4}
V_d=\frac{4 \pi}{3} R^3 f_{V T} (\Theta); \quad f_{V T} (\Theta) \equiv
\frac{1}{4}(2 + \cos \Theta)(1-\cos \Theta)^2,
\end{equation}
\begin{equation} \label{eq5}
A= 2 \pi R^2 (1-\cos \Theta),
\end{equation}
and
\begin{equation} \label{neqq1}
A_b=R^2 \pi \sin^2 \Theta,
\end{equation}
where $f_{VT}$ is the Volmer-Turnbull (VT) function \cite{1,41}.
The first term on the right hand side of Eq. (\ref{eq6}) can be motivated by considering that
the volume contribution to $\Delta F$ is typically written as $-V_d \Delta p$, where $\Delta p$
is the pressure difference between the coexisting phases. The linear expansion of $\Delta p$,
by recognizing that density is a derivative of pressure with respect to the chemical
potential, provides this form.
A further assumption here is that the liquid and vapor densities, for the
overall box density $\rho_{\rm box}$, are very close to the
corresponding coexistence densities in the thermodynamic limit.

In this paper, we shall ignore the possibility that the vapor-liquid interfacial
tension $\gamma_{\ell \upsilon}$ contains a curvature correction \cite{8,32,33,34,35,36,37,38,39,40} of
order $1/R$, which would make the contribution of the term related to the upper
surface area in Eq. (\ref{eq6}) of the same order $(\propto R)$ as
the line tension term \cite{10} (when $R$ dependence in $\tau$ is ignored).
In fact, for the models that we shall
explicitly study, viz., the Ising lattice gas and a symmetrical binary Lennard-Jones fluid, no such
curvature correction is expected -- the leading correction there is proportional \cite{36,37,44,44pr} to
$R^{-2}$. Here we shall assume throughout that the radii of
interest are large enough such that this curvature dependence of
$\gamma_{\ell \upsilon}$ \cite{8,32,33,34,35,36,37,38,39,40} can be neglected. This fact
we will justify later.
For the lattice gas, however, it is a severe approximation
to ignore the change in interfacial free energies due to lattice
anisotropy (see e.g.~Ref. \cite{23}
for a discussion). To avoid this problem, 
one should, in fact, choose for the latter model the vicinity
of the critical point. Critical-like fluctuations then will get rid of the anisotropy, allowing one 
to have spherical droplets in the bulk and sphere-cap
shaped droplets on planar walls. As a passing remark, for
$\gamma_{wl} (\delta\mu)$ and $\gamma_{wv} (\delta\mu)$ we do not discard here the possibility of the
leading correction being proportional to $1/R$. In fact, we will see that
$\delta\mu \sim 1/R$ which brings a correction linear in $1/R$ to the
wall tensions.

When we consider a large spherical droplet in the bulk, we have the simpler result [cf. Eq. (\ref{eq6})]
\begin{equation} \label{eq7}
\Delta F_{\rm sphere} =-\frac{4 \pi}{3} R^3 \delta \mu (\rho_\ell^{\rm coex} - \rho_v^{\rm coex}) + 4 \pi R^2 \gamma_{\ell v}.
\end{equation}
Extremizing this expression, viz., by equating $\partial \Delta F_{\rm sphere} /\partial R \mid_{_{R^*}}$
to zero, one obtains the standard theoretical expressions for the critical radius $R^*$ and 
associated nucleation free energy barrier $\Delta F^*_{\rm hom}$ related 
to the homogeneous nucleation \cite{1,2,3,4}.
These are
\begin{equation} \label{eq8}
R^* = \frac{2 \gamma_{\ell v}}{\delta \mu(\rho^{\rm coex}_\ell - \rho^{\rm coex}_v)},~ \Delta F^*_{\rm hom} =\frac{4 \pi}{3} R^{*^2}  \gamma_{\ell v}.
\end{equation}
When the contribution of the line tension to the droplet free energy is negligible,
the standard result for the barrier, $\Delta F^*_{\rm het}$, against heterogeneous
nucleation would result \cite{1,41}. If we assume that the contact angle takes its
macroscopic value $\Theta_\infty$, then
\begin{equation} \label{eq9}
\Delta F^*_{\rm het} =\Delta F^*_{\rm hom} f_{V T} (\Theta_\infty) =\frac{4 \pi}{3} R^{*^2} \gamma_{\ell v} \, f_{VT} (\Theta_\infty),
\end{equation}
the same VT function $f_{VT}(\Theta_\infty)$ [see Eq.~(\ref{eq4})], as in the case of volume,
describing the reduction of the barrier in comparison with the homogeneous case.

In our previous simulation studies, attempting to extract the line tension
\cite{18,19,20}, the $\delta \mu$-dependence of the wall excess free energies as well as
the change of the contact angle due to line tension [see Eqs.~(\ref{eq2}) and (\ref{eq3})] or
other reasons, were neglected. Thus, Eq.~(\ref{eq9}) was simply replaced by an
expression
\begin{equation} \label{eq10}
\Delta F^{*\infty}_{\rm het} =\frac {4 \pi}{3} R^{*^2} \gamma_{\ell \upsilon} f_{VT} (\Theta_\infty) + 2 \pi R^* \tau \sin \Theta_\infty,
\end{equation}
that is believed to be correct only in the limit $R^* \rightarrow \infty$.
However, taking Eq.~(\ref{eq3}) into account, one rather finds \cite{15}, minimizing the
free energy at constant droplet volume, that
  \begin{equation} \label{eq11}
\Delta F^{*}_{\rm het} =\frac{4 \pi}{3} R^{*^2} \gamma_{\ell \upsilon} f_{VT} (\Theta) + \pi R^* \tau \sin \Theta.
\end{equation}

Note that part of the correction due to the line tension is accounted for by
replacing $\Theta_\infty$ by $\Theta$ in $f_{VT}$. This is seen by expanding
Eq.~(\ref{eq3}) in leading order in $\Theta-\Theta_\infty$ and
considering the first term on the right hand side of Eq.~(\ref{eq11}):
\begin{equation}\label{neqn1}
\frac{\tau}{R \gamma_{\ell \upsilon}} \simeq \sin^2 \Theta_\infty (\Theta - \Theta_\infty),
\end{equation}
and
\begin{equation}\label{neqn2}
\frac{4 \pi}{3} R^{*^2} \gamma_{\ell \upsilon} f_{VT} (\Theta) \simeq \frac{4 \pi}{3} R^{*^2} \gamma_{\ell \upsilon} f_{VT} (\Theta_\infty) + \pi R^* \tau \sin \Theta_\infty.
\end{equation}
Accepting Eq.~(\ref{eq11}), one can show that for large $R^*$ the error made
by Eq.~(\ref{eq10}) scales like $\tau^2$:
\begin{equation} \label{eq12}
\Delta F^*_{\rm het} - \Delta F^{*\infty}_{\rm het} \simeq \frac{\pi \cos \Theta_\infty}{\gamma_{\ell \upsilon}
\sin^2 \Theta_\infty} \tau^2.
\end{equation}
Clearly, it is desirable to avoid Eq.~(\ref{eq10}), which
for small $R^*$, as used in previous simulations, is unjustified.

One purpose of the present work
hence is to reconsider the simulations presented in Refs. \cite{18,19,20} and attempt an
analysis where both $\tau$ and $\Theta$ are extracted from the simulation data by avoiding the
use of Eq.~(\ref{eq10}) completely.
Ref. \cite{15} contains a detailed derivation of Eq.~(\ref{eq11}), promising ``a
rigorous thermodynamic formulation''. However, this derivation also ignores,
to some extent like that of Eq. (\ref{eq10}),  both a
possible dependence of $\tau$ on $R$ and $\Theta$, and the possible
$\delta \mu$-dependence of the wall tensions, by the step from Eq.~(\ref{eq2}) to
Eq.~(\ref{eq3}). In view of these approximations, in our proposed approach we keep
in mind that the estimates should be done in such a way that it does not matter to
what extent the difference $\Theta-\Theta_\infty$ is due to the line tension or due
to the possible $\delta \mu$-dependence of the wall tensions. 

A second purpose of the present work is to
demonstrate that from a simulation of wall-attached droplets in equilibrium in finite systems in the
canonical ensemble, as indicated in Fig.~\ref{fig1}, one can obtain $\Delta F^*_{\rm het}$ directly, as a
function of $\delta\mu$, without the need to use the above equations. The idea is to study the density excess
$\rho_{\rm box} - \rho_v$ 
(see Fig.~\ref{fig1}b) and the associated free energy difference between the two systems
(with and without droplet) at the same $\delta \mu$.

\section{Ising model simulations and their analysis}

We study the nearest-neighbor Ising model, on a simple cubic lattice, at a temperature
$k_BT/J=3.0$, $J$ being the exchange constant and $k_B$ the Boltzmann 
constant. At this temperature, effects due
to critical fluctuations are still negligible, given that the critical temperature is \cite{42}
$k_BT_c/J\simeq 4.51$. At the same time, the temperature is high enough
so that the anisotropy effects on the interface are
sufficiently small and thus, the droplet shape is almost
spherical \cite{43}.

\begin{figure} [ht]
\subfigure[]{
\includegraphics[width=4.5cm,clip]{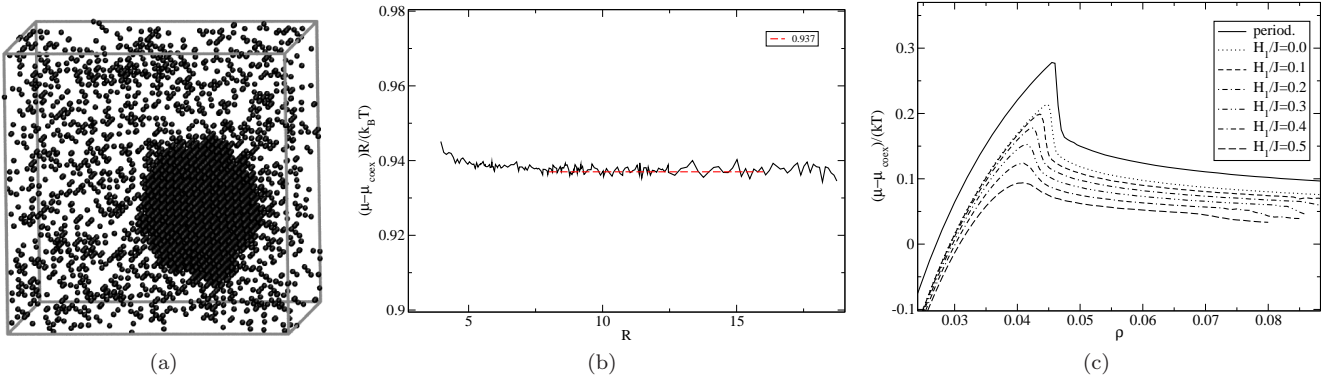}
}
\subfigure[]{
\includegraphics[width=6.5cm,clip]{fig3b.eps}
}
\subfigure[]{
\includegraphics[width=6cm,clip]{fig3c.eps}
}
\centering
\caption{\label{fig3} (a) Snapshot of an Ising model in the bulk
at $k_BT/J=3.0$ in a $L\times L \times L$ simulation box with $L=40$ lattice
units, periodic boundary conditions throughout, and $\rho_{\rm box}=0.1$. Occupied lattice sites are marked by dots.
(b) $\delta\mu R/k_BT$ is plotted vs. $R$, for $k_BT/J=3.0$. Note that for $R \geq 7$
there is no systematic $R$-dependence of this product any more. The observed average
value (0.937(1)), highlighted by a broken horizontal line, 
slightly exceeds the theoretical value 0.889(1), Eq.~(\ref{eq8}),
predicted from the interface tension $\gamma_{\ell \upsilon} /k_BT=0.434(1)$ for
planar interfaces, due to slight enhancement of the effective interface tension
\cite{43} resulting from the slightly non-spherical droplet shape.
Here $\rho^{\rm coex}_\ell -\rho^{\rm coex}_v=0.948$ was also used.
(c) Chemical potential difference $\delta\mu/k_BT$ is plotted
vs. $\rho$ at $k_BT/J=3.0$, for a $L \times L \times L$ system with periodic
conditions and $L \times L \times L_z$ systems [of Fig.~\ref{fig1} (a)] with different
choices of the boundary field $H_1/J$, as indicated. Here $L=L_z=40$
throughout. } \end{figure}

As depicted in Fig. \ref{fig1} (a), we choose a geometry
restricted in the $z$-direction with linear dimension $L_z$. At the bottom layer ($n=1$),
perpendicular to the latter direction, a short-range ($\delta$-function) positive 
surface field $H_1$ acts. An
antisymmetric boundary condition is created by putting a surface field of
equal magnitude, but of opposite sign, at $n=L_z$. On the other hand, periodic
boundary conditions are applied in $x$ and $y$ directions, for which the linear dimensions 
$L_x$ and $L_y$ equal $L$. We emphasize that the present approach, as described below,
does not rely on a ``microscopic'' identification
of which spins do belong to a droplet or not. While such an identification would
be possible \cite{43}, the strong fluctuations in the interfacial region
[see Fig.~\ref{fig3} (a) for a picture in a bulk system]
make such microscopic approach inconvenient. Before getting
into the situation with a wall-attached droplet we provide a brief discussion with
respect to the bulk.

In this work we employ a lattice variant of the Widom particle insertion method \cite{R1}, which is used
to determine the chemical potential as a function of density [Fig.~\ref{fig3}(c)]. 
This way nuclei of various shapes and
sizes in a system can be probed. Integration of this curve with respect to density yields the free energy.
Alternatively, one could employ a successive umbrella sampling technique \cite{R2,R3} by varying the overall
composition in the system in a quasistatic manner \cite{18,19}. The method allows one to obtain the probability distribution
for the density in the full range of the latter. From there one could obtain a free energy profile that
contains information on the bulk coexistence values of densities as well as excess free energies
for interfaces and lines. The chemical potential difference, $\delta\mu$,
for a particular structure and size
with respect to the bulk coexistence, could then be obtained from this free
energy profile, by taking derivative with respect to density.

The basic principle one needs to apply then is that a liquid
droplet of density $\rho_{\ell}$ coexists with a vapor of density $\rho_v$ at the
same value of $\delta\mu$, as depicted in Fig. \ref{fig1} (b). This, for a
particular overall particle density, $\rho_{\rm box}$, inside the box, identifies
$\rho_{\ell}$ and $\rho_v$. The latter allows one to estimate the volume contribution
of free energy, subtraction of which from the total value provides the excess free
energy at $\rho_{\rm box}$. In addition, the corresponding
radius $R$, in a bulk system of volume $V_{\rm box}$, can be extracted from the lever
rule as
\begin{equation} \label{eq13}
\rho_{\rm box} =\rho_v + (\rho_\ell - \rho_v) \frac{4 \pi R^3}{3 V_{\rm box}},
\end{equation}
for a spherical structure of the liquid droplet.
This, of course, implicitly implies the use of an equimolar dividing surface between vapor
and liquid so that there are no excess particles attributed to the interface.
Since, for any choice of $\rho_{\rm box}$ we
know the appropriate value of $\delta \mu$ and $R$ is
straightforwardly calculated, the relation $\delta \mu R$ vs. $R$ can be constructed.
The corresponding result in Fig.~\ref{fig3} (b), from bulk, implies an inverse relation
between $\delta\mu$ and $R$ for large enough $R$. We make use of this relation in the situation
with antisymmetric walls as well.
In the confined case, with surface
fields $\pm H_1$, we have
\begin{equation} \label{eq14}
\rho_{\rm box} =\rho_v + (\rho_\ell - \rho_v) \frac{4 \pi R^3 f_{VT} (\Theta)}{3 V_{\rm box}}.
\end{equation}

Of course, a crucial assumption of this approach is that also for radii $R$ that are
finite, but much larger than the lattice spacing, the deviation of the shape of a
droplet from a sphere-cap can be neglected. Only when this assumption is accurate,
one can write the volume of the wall-attached droplet as \cite{1,41} $4 \pi R^3 f_{VT} (\Theta)/3$.
Assuming that Eq.~(\ref{eq14}) is accurate, for any given choice of $\rho_{\rm box}$
we know $\delta \mu$  [cf.~Fig.~\ref{fig3} (c), for simulation results from confined systems],
as well as the radius $R$ from bulk simulation
[cf.~Fig.~\ref{fig3} (b)]. Since at the obtained  value of $\delta \mu$, $\rho_v$ and
$\rho_\ell$ are also known, we can use this information to obtain
$f_{VT} (\Theta)$ from Eq.~(\ref{eq14}). This allows us to calculate
hence the contact angle $\Theta$ from Eq.~(\ref{eq4}). E.g.,
for the case $H_1=0$ and $R=10$, this analysis yields $\Theta \approx 87.3^o$, while
$\Theta_\infty =90^o$. Although this difference $\Theta_\infty - \Theta$ is relatively
small, it must not be neglected, contrary to the previous studies \cite{18,19,20}.
We emphasize that for this method of estimation of the contact angle of wall-attached sphere-cap
shaped droplets for a chosen value of $\delta \mu$ (or $R$) from
Eq.~(\ref{eq14}), at a chosen value of $H_1/J$, neither the knowledge of $\Theta_\infty$
nor the knowledge of the line tension $\tau$ is required.

At this point we mention that the assumptions implied by Eq.~(\ref{eq14}) can be avoided if we only
like to know the excess number of particles 
\begin{equation}\label{nex}
N_{\rm exc}= L^2 L_z (\rho_{\rm box} - \rho_v),
\end{equation} 
due to
the wall-attached droplet in the simulation box. This is of interest, since the ``volume term'' of nucleation
theory can also be written as $ N_{\rm exc} \delta \mu$. We will make use of this later.

\begin{figure} [ht]
\includegraphics[width=8cm,clip]{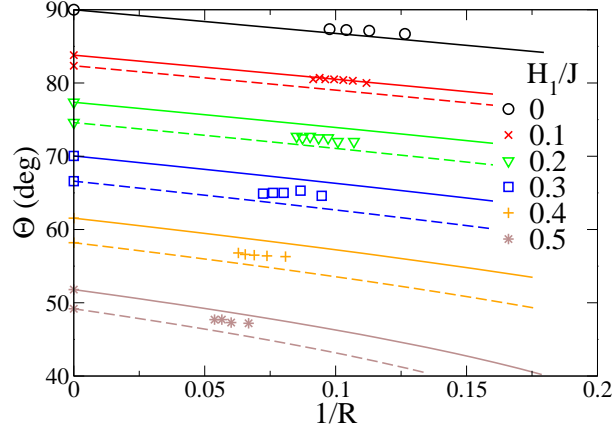}
\centering
\caption{\label{fig4}Plot of the contact angle $\Theta$ of wall-attached droplets,
for several choices of $H_1/J$ (symbols), against $1/R$, obtained by assuming a sphere-cap shape
of the droplet and using the data of Fig.~\ref{fig3}(c) and Eq.~(\ref{eq14}).
The curves were constructed by using the line tension estimates for 
planar inclined interfaces (extracted in Ref. \cite{23}) in Eq. (\ref{eq3}).
The broken curves correspond to $\Theta_{\infty}$ obtained from the Young's equation and
the full curves used the values of $\Theta_{\infty}$ that were estimated by accounting for the
anisotropy effects on the interfacial free energy (see Ref. \cite{23}).}
\end{figure}

Fig.~\ref{fig4} now shows some central results obtained in the present paper, viz.,
estimates for $\Theta$ as a function of $1/R$, for various choices of $H_1/J$, as indicated
in the figure. Note that the values $\Theta_\infty$, shown for $1/R=0$, are from the study of
Block et al. \cite{23}, for planar inclined interfaces in the Ising model.
For $H_1 > 0$ two estimates of $\Theta_\infty$ are included: the smaller values are obtained
from Young's equation, Eq.~(\ref{eq1}), while the upper ones are
from a ``first principle's'' approach \cite{23}. The latter method
takes the effects of lattice anisotropy on the
interfacial free energy, $\gamma_{\ell \upsilon}$, of the lattice gas model, into account.
This lattice effect is expected to become negligible when the temperature approaches the
critical value, but turns out to yield a difference of a few degrees for
the contact angles at the considered temperature $k_BT/J=3.0$. In any case, it is remarkable
to observe that in the accessible range of radii $(8 \leq R \leq 20)$ the contact angles
of the droplets are always smaller than both the estimates for $\Theta_\infty$. Roughly,
the variation of $\Theta$ with $1/R$ is compatible with a linear behavior, as expected
from Eq.~(\ref{eq3}), when $R \rightarrow \infty$ [see Eq. (\ref{neqn1})].

As a first qualitative test of Eq.~(\ref{eq3}), in Fig. \ref{fig4} we plot the result of this equation,
using for $\gamma_{\ell \upsilon}$ the estimate
$\gamma_{\ell \upsilon} (R \rightarrow \infty)/k_BT =0.444$,
and for $\tau(\Theta)$ the estimates due to Block et al. \cite{23}
for planar inclined interfaces that are hitting a flat wall. It is seen that
the resulting curves that use $\Theta_\infty$ from Eq.~(\ref{eq1}) always fall slightly
below the actual data for $\Theta$, while the other predictions on $\Theta_\infty$,
taking the anisotropic feature of the interfacial tension into account, overestimate it. Clearly,
the effects due to this anisotropy, which also is expected to cause a slight but
systematic deviation from sphere shape \cite{43}, preclude a more precise test of Eq.~(\ref{eq3}).

Next we consider the numerical computation of the excess free energy of the droplet,
making use of a thermodynamic integration  procedure
\begin{equation} \label{eq15}
\Delta f_s (\rho_{\rm box} ) =\int\limits_{\rho_v}^{\rho_{\rm box}} \delta \mu (\rho) d \rho,
\end{equation}
using data such as shown in Fig.~\ref{fig3} (c). Eq.~(\ref{eq15}) is the excess free
energy due to the droplet per lattice site, in the canonical ensemble. Note that
Eqs.~(\ref{eq6}) and (\ref{eq7}) refer to the grand-canonical ensemble. In the canonical
ensemble the volume term proportional to $\delta \mu$ is not present. So we pick up
from $\Delta f_s(\rho_{\rm coex})$ just the surface and line tension contributions.
Since there is a one-to-one correspondence between $\rho_{\rm box}$ and $\delta \mu$
and hence $R$, the quantity $L^2 L_z \Delta f_s(\rho_{\rm box})$ can be
re-interpreted in terms of the surface excess free energy $F_s(R)$. Recently, it has
been recognized that the translational entropy of the droplet in the simulation box
should not be ignored \cite{40}, if one compares data obtained from simulations
[Eq.~(\ref{eq15})] with theoretical estimates. For a wall attached droplet, we hence
should add $k_BT \ln (L^2)$ to  $\Delta f_s(\rho_{\rm box})$ to obtain $F_s(R)$. For $L=40$ 
this amounts to about $7.38 k_BT$.

In previous work it was assumed that $F_s(R)$ can be interpreted as \cite{18,19}
\begin{equation} \label{eq16}
F_s(R)=4 \pi R^2 \gamma_{\ell \upsilon} \, f_{VT} (\Theta_\infty)´+ 2 \pi R \tau  \sin \Theta_\infty,
\end{equation}
which leads to the barrier for heterogeneous nucleation -- see Eq.~(\ref{eq10}).
For an understanding on the effects of curvature dependence of $\gamma_{\ell v}$,
by neglecting the correction due to the line tension, the surface part of the
free energy of the wall-attached droplet simply could be written as
\begin{equation}\label{nneqn1}
F_s(R)=4 \pi R^2 f_{VT} (\Theta) \gamma_{\ell \upsilon}(R).
\end{equation}
At $k_BT/J=3.0$, the data of Winter et al. \cite{18,19} are compatible
with
\begin{equation}\label{nneqn2}
\gamma_{\ell \upsilon} (R)\simeq \frac{\gamma_{\ell \upsilon} (\infty)}{1+(a/R)^2},
\end{equation}
with
$\gamma_{\ell \upsilon} (\infty)/k_BT=0.444$ and $a$=1.85 lattice units. Note that for
$R \geq 8$ this result implies for the total free energy of a droplet in the bulk
to be [cf. Eq. (\ref{eq7})]
\begin{equation} \label{eq17}
\Delta F_{\rm sphere} (R)=-\frac{4}{3} \pi R^3 (\rho^{\rm coex}_\ell - \rho^{\rm coex}_ \upsilon) \delta \mu + 4 \pi R^2 \gamma_{\ell \upsilon} (\infty) - 4 \pi \gamma_{\ell \upsilon} (\infty) a^2.
\end{equation}
Hence the result for $R^*$ [see Eq.~(\ref{eq8})], resulting from
$\partial [\Delta F_{\rm sphere} (R)]/\partial R] \mid_{R^*} =0$, is not truly affected by
this correction due to the $R$-dependence of $\gamma_{\ell \upsilon}(R)$.

\begin{figure} [ht]
\includegraphics[width=8cm,clip]{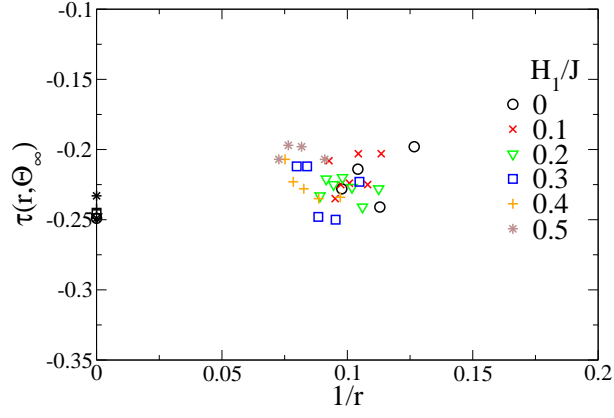}
\centering
\caption{\label{fig5} Plots of the line tension $\tau$, in units of $k_BT$, vs.
the inverse circular radius $1/r$, for various choices of $H_1/J$, as indicated, for
$k_BT/J=3.0$, in the simple cubic Ising model. Data for $r^{-1}=0$ were obtained for
planar interfaces by Block et al. \cite{23}.}
\end{figure}

Naively one might correct Eq.~(\ref{eq16}), by rewriting it after taking the actual contact
angles into account, as
\begin{equation}\label{nneqn3}
F_s(R)=4  \pi R^2 f_{VT} (\Theta) \gamma_{\ell \upsilon} (R) + 2 \pi  r \tau (r, \Theta),
\end{equation}
where $r=R \sin \Theta$ is the radius of the circular contact line. The resulting
estimates for $\tau (r, \Theta)$ are plotted in Fig.~\ref{fig5}, versus
$1/r$. Unfortunately, the scatter of the resulting data is large and not systematic,
indicating that the statistical accuracy with which $F_s(R)$ is obtained from
Eq.~(\ref{eq15}), as well as the accuracy of $\Theta$, does not suffice for a
meaningful analysis of the dependence of the line tension on the circular radius
$r$ and contact angle $\Theta$. It is seen that the results for a planar interface
(from Block et al. \cite{23}) increase slightly from $\tau(\infty, 90^o)\approx-0.249$ to
$\tau (\infty, 52^o) \approx -0.233$, when $\Theta_\infty$ is varied. The data for $\tau(r, \Theta)$
extracted from the droplets seem to be only very roughly consistent with this trend,
but probably lack the necessary accuracy to allow for a clear conclusion. But it is evident
that the dependence of $\tau$ on $R$ and $\Theta$ is not very strong.
However, if Eq.~(\ref{eq11}) would be used to analyse the data for $F_s(R)$, the
absolute magnitude of $\tau (r, \Theta)$ would be twice as large, making it completely
incompatible with the limiting behavior for $r \rightarrow \infty$ (planar interface).

At first sight, all these facts are surprising.
However, while the droplet surface  is a two-dimensional object here, the contact
line is one-dimensional, and hence much larger finite-size corrections due to
fluctuations can be expected. When, for the sake of analogy, we consider the droplets
in the two-dimensional Ising model, where the droplet ``surface'' also is a
one-dimensional line, a much larger size effect than in $d=3$ [see Eq. (\ref{nneqn2})] 
occurs \cite{34,40}
\begin{equation} \label{eq18}
\gamma_{\ell \upsilon} (R)=\gamma_{\ell \upsilon} (\infty) + \frac{5}{4 \pi} \,\frac{\ln R}{R} + \frac{\rm const}{R}.
\end{equation}

Returning now to the starting point of our discussion, introducing the line tension
from the mechanical equilibrium of the contact line, Eq.~(\ref{eq2}), we note that
the wall tensions, to first order in $\delta \mu$, can be written as
\begin{equation}\label{nneqn4}
\gamma_{wv} (\delta \mu)=\gamma_{wv} (0) + \delta \mu \Gamma_v,
\end{equation}
and
\begin{equation}\label{nneqn5}
\gamma_{w\ell} (\delta \mu)=\gamma_{w\ell} (0) + \delta \mu \Gamma_{\ell},
\end{equation}
which yield [see Eqs. (\ref{eq1}) and (\ref{eq2})]
\begin{equation} \label{eq19}
\gamma_{\ell \upsilon} (\cos \Theta-\cos \Theta_\infty)=\delta \mu \Delta \Gamma - \frac{\tau}{R \sin \Theta},
\end{equation}
where $\Delta \Gamma$ is the difference in the adsorption from the vapor
$(\Gamma_v)$ and the liquid $(\Gamma_\ell)$.
Using Eq.~(\ref{eq8}), to replace $\delta \mu$ by $R$, we note that the first term on
the right hand side of Eq.~(\ref{eq19}) is of the same order as the second term.
Still another expression has been derived in Ref. \cite{10} $(r=R \sin \Theta)$:
\begin{equation} \label{eq20}
\gamma_{\ell \upsilon} \Big(\cos \Theta-\cos \Theta_\infty \Big) = -\frac{\tau}{R \sin \Theta} -\frac{d \tau}{dr}-\frac{1}{R} \cos \Theta_ \infty \frac{d \tau}{d \Theta},
\end{equation}
where a further correction involving the Tolman length \cite{32} has been omitted.
Eq.~(\ref{eq2}) [or Eq. (\ref{eq19})] is compatible with Eq.~(\ref{eq20}) only when $\tau (r, \Theta)$
depends neither on $r$ nor on $\Theta$, which clearly is not true.

\begin{figure} [ht]
\includegraphics[width=8cm,clip]{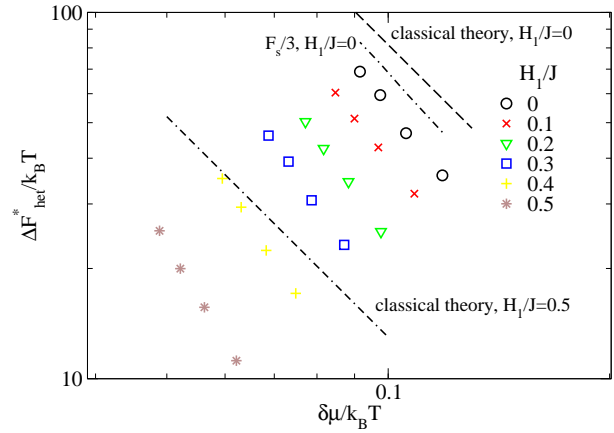}
\centering
\caption{\label{fig6} Nucleation barrier for $k_{B}T/J=3.0$ $(L=40)$ in the simple cubic 
Ising model as a function of chemical potential $\delta\mu/k_{B}T$. The simulation results
for various values of $H_1/J$ are represented by symbols. For $H_1/J=0$ and $0.5$ these
are compared with the classical theory (curves). For $H_1/J=0$ we have also shown a plot of $F_s/3$
(see text for details).
}\end{figure}

While our work hence cannot fully clarify the problems relating to the line
tension, we do get from our study valid and reasonably accurate results for the barrier against heterogeneous nucleation as
\begin{equation}
\frac{\Delta F^*_{\rm het}}{k_BT} =-\frac{\delta \mu}{k_BT} N_{\rm exc} + F_s,
\end{equation}
where $F_s=\Delta f_s(\rho_{\rm box}) + k_BT \ln (L^2)$, as noted above. Fig.~\ref{fig6}
shows a log-log plot of $\Delta F^*_{\rm het}/k_BT$ versus $\delta \mu/k_BT$. For $H_1/J=0$ and $0.5$
we include a comparison with the prediction of the classical theory [see Eqs. (\ref{eq8}) and (\ref{eq9})], 
\begin{equation}\label{neweq1}
\frac{\Delta F^*_{\rm het, class}}{k_BT}=
\frac{4 \pi}{3} \frac{R^{*2} \gamma_{\ell v}f_{VT}(\Theta_{\infty})}{k_BT} = 
\frac{16 \pi}{3}\Big(\frac{\gamma_{\ell v}}{k_BT}\Big)^3 (\rho^{\rm coex}_\ell - \rho^{\rm coex}_v)^{-2}\Big(\frac{\delta\mu}{k_{B}T}\Big)^{-2} f_{VT}(\Theta_{\infty}).
\end{equation} 
One sees that the actual barriers, say for $H_1/J=0$, are about a factor 1.6 lower, 
though the general trend is similar. If there is no line tension
effect, one would also have the relation $\Delta F^*_{\rm het}=F_s/3$.
For $H_1/J=0$ this estimate also has been
included, with the above choice of $F_s$, and one sees that this still amounts to an overestimation. Similar, but somewhat larger,
discrepancies apply to the other choices of $H_1/J$ as well. 
In fact the simulation results for $\Delta F^*_{\rm het}$ are expected to be smaller than
the predictions. This is because the negative values of line tension (see Fig.~\ref{fig5}) 
reduce the barriers.

Unfortunately, for each choice of $H_1/J$ only a rather restricted range of $\delta \mu/k_BT$ is accessible: this happens because we cannot use
data for $\rho=\rho_{\rm box}$ [in Figs.~\ref{fig1}(b) and \ref{fig3}(c)] that are close to the peaks in the $\delta \mu$ vs. $\rho$ curve; as is well known,
these peaks correspond to the droplet evaporation/condensation transition \cite{18,19,36}. Likewise, data for too large $\rho$ (where in the
curves of Fig.~\ref{fig3}(c) an inflection point is seen) cannot be used either; there the droplet shape changes from sphere cap to cylinder cap
shape (stabilized by the periodic boundary conditions). But the range of barriers that is accessible here (about $10 < \Delta F^*_{\rm het}/k_BT < 60)$
is in a range that would be physically significant. Note that the regime of much larger droplets, for which the theory outlined in Sec. II is presumably
more accurate, would relate to much larger barriers which are physically irrelevant.

\section{Analysis of simulation results for the symmetric binary fluid}

The inter-particle interaction in the binary ($A,B$) fluid mixture, in this section,
is defined in terms of the Lennard-Jones (LJ) potential ($r=|\vec{r}_i-\vec{r}_j|$,
$\vec{r}_i$ and $\vec{r}_j$ being the positions of particles $i$ and $j$, respectively)

\begin{equation} \label{eq21}
U_{LJ} = 4 \varepsilon_{\alpha \beta}[\big(\frac{\sigma_{\alpha \beta}}{r}\big)^{12} -
\big(\frac{\sigma_{\alpha \beta}}{r}\big)^6], \quad \alpha, \beta \in A, B,
\end{equation}
with
\begin{equation} \label{eq22}
\sigma_{AA}=\sigma_{BB}=\sigma_{AB}=\sigma, \quad \varepsilon_{AA}=\varepsilon_{BB} =\varepsilon.
\end{equation}
For the interaction strength, $\varepsilon_{AB}$ is chosen differently \cite{20,44,45}, viz.,
$\varepsilon_{AB} =\varepsilon/2$, to facilitate liquid-liquid phase separation,
as opposed to the choice of equal diameter ($\sigma$)
for all combinations of particles. To improve the speed of the simulations, it is preferable to cut
the potential at some distance $r=r_c$, for which we choose $2.5\sigma$. But for molecular dynamics
simulations (that were performed \cite{45,48}
for understanding of various equilibrium and nonequilibrium dynamical
properties of this model) it became essential to make the potential and also the force
continuous at $r=r_c$. This can be  achieved by modifying the LJ potential as \cite{20,44,45}
\begin{equation} \label{eq23}
u(r)=U_{LJ} (r)-U_{LJ} (r_c) - (r-r_c) \frac{d U_{LJ} (r)}{dr} \mid_{_{r=r_c}},
\end{equation}
which we used for the present study.

Temperature in this model has the unit $k_B/\varepsilon$,
all lengths will be measured in unit of $\sigma$ and the
dimensionless density is calculated as $\rho_{\rm box}=N\sigma^3/V_{\rm box}$.
Here $N$ is the total number of particles ($=N_A+N_B$, $N_A$ and $N_B$ being the numbers for $A$
and $B$ particles) and $V_{\rm box}=L_xL_yL_z \sigma^3$,
where $L_{\alpha}$ ($\alpha = x,~y,~z$) is the box length
along the Cartesian direction $\alpha$. In the following, for convenience, we set $k_B$, $\varepsilon$ and
$\sigma$ to unity. We may, however, explicitly use them when verification of an expression with
respect to dimension is required. We set $\rho_{\rm box}=1$
and perform all simulations at $T=1$, which is far
below the critical temperature ($T_c\simeq 1.421$) of the model \cite{45,48}. The bulk interfacial tension
for this temperature was found to be \cite{20} $\gamma_{AB}=0.722\pm 0.002$.
This model shares with the Ising model the advantage of a strict symmetry between the two
coexisting phases, thus, the critical composition is identically set at $50:50$
composition of $A$ and $B$ particles. The model has the further advantage that the
interfacial tension is fully isotropic, due to the off-lattice character.

In order to realize a situation with antisymmetric walls, similar to Fig.~\ref{fig1}(a),
wall potentials $u_A (z)$ and $u_B(z)$, acting on the two types of particles, are chosen as \cite{20,49,44}
\begin{equation} \label{eq24}
u_A(z) =\frac{2 \pi \rho}{3} \Big[\varepsilon_r \Big\{\Big(\frac{\sigma}{z + \delta}\Big)^9 + \Big(\frac{\sigma}{L_z +\delta -z}\Big)^9 \Big\} - \varepsilon_a \Big(\frac{\sigma}{z + \delta}\Big)^3 \Big],
\end{equation}
\begin{equation} \label{eq25}
u_B(z) =\frac{2 \pi \rho}{3} \Big[\varepsilon_r \Big\{\Big(\frac{\sigma}{z + \delta}\Big)^9 + \Big(\frac{\sigma}{L_z +\delta -z}\Big)^9 \Big\} - \varepsilon_a \Big(\frac{\sigma}{L_z + \delta-z}\Big)^3 \Big],
\end{equation}
where $z$ ($0 \leq z \leq L_z$), as in the previous section,
again is the coordinate perpendicular to the walls.
An offset $\delta =\sigma/2$, for the origin of the wall potentials, is used, such that the
latter is finite everywhere inside the box. Both walls exert the same repulsive potential
(with $\varepsilon_r=\varepsilon/15$) on both types of particles, so that no particles can
leave the system by penetrating the walls.
Note that periodic boundary conditions are used in the $x$ and $y$ directions, with $L_x=L_y=L$.
The attractive potential, with prefactor $\varepsilon_a$ [see the last terms on the
right hand sides of Eq. (\ref{eq24}) and (\ref{eq25})], acts on $A$ particles only from the wall at $z=0$,
while for $B$ particles it comes
only from the wall at $z=L_z$. This is the analog of the antisymmetric surface field $\pm |H_1|$
in the Ising model (Fig.~\ref{fig1}). However, while for the lattice model the wall
potential was chosen to be of strictly short range, here we do not use any cut-off for the wall
potentials.

\begin{figure} [ht]
\includegraphics[width=8cm,clip]{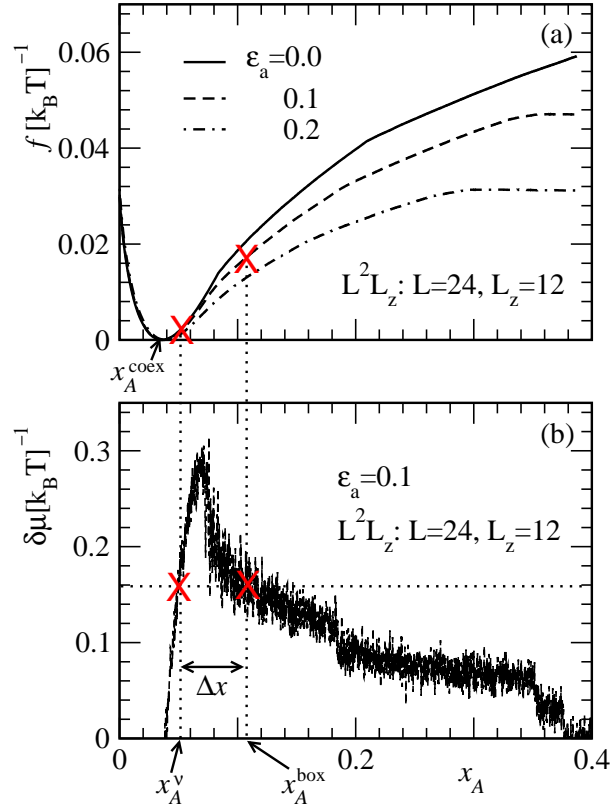}
\centering
\caption{\label{fig7} (a) Plots of the effective free energy $f(x_A, T)$ for
three values of $\varepsilon_a$, as quoted in the figure, versus $x_A$, the concentration
of $A$ particles during umbrella sampling Monte Carlo simulations 
of the symmetric binary fluid. The chosen linear dimensions
are $L=24$ and $L_z=12$. Compositions and values of $f$ for two states (one with and the other without 
a droplet -- see below) at the same chemical potential difference,
identified by a dotted horizontal straight line [see part (b)], for $\varepsilon_a=0.1$,
are marked by crosses.
(b) Chemical potential difference $\delta \mu$, in units of $k_BT$, is plotted vs.
$x_A$, for the case $\varepsilon_a=0.1$, $L=24$, and $L_z=12$. The
state with $\Delta \mu=0$ corresponds to bulk coexistence, $x_A^{\rm coex}$.
State $x^\nu_A$ [marked with a cross on the left] 
is the analog of the state denoted as $\rho_v$ in
Fig.~\ref{fig1}(b) for the vapor-liquid coexistence, and the state $x^{\rm box}_A$ is the
chosen average concentration where an A-rich droplet coexists with the surrounding B-rich
fluid at the same chemical potential difference.
} \end{figure}

Our Monte Carlo (MC) simulations \cite{42} with this model, of course, consider local displacement moves of
the particles. For such trial moves we randomly choose particles and  change
their Cartesian coordinates, again randomly, such that the displacement lies in the range
$[-\sigma/20, \, +\sigma/20]$. When $N$, $V_{\rm box}$ and $T$ are kept fixed, this 
provides realization of the canonical ensemble \cite{42}.
For the purpose of efficiently obtaining
the effective free-energy density $f(x_A, T)/k_BT$, $x_A$ ($=N_A/N$),
the variable analogous to $\rho$ for the vapor-liquid case in the previous section, being the
concentration of $A$ particles, in the two-phase
coexistence region, we use (successive) umbrella sampling method
\cite{20,R2}. Thus, in addition to using displacement as trial moves, we have
used identity switches ($A\rightarrow B\rightarrow A$) as well. For such moves we have chosen a
particle randomly, identified its type and changed it. This calls for an
introduction of difference between chemical potential between $A$ and $B$ particles
in the Boltzmann factor, for the execution of the Metropolis algorithm \cite{42}. However, along
coexistence such difference is identically zero. Umbrella sampling MC simulations,
in addition to providing information on the coexistence curve, helps obtaining the accurate
probability distribution for the fluctuation of $x_A$
over the whole spectrum ($\in [0,1]$) of
the latter. From such probability distribution the free energy curve, as a function of $x_A$,
can be straightforwardly obtained, for bulk as well as for the confined systems.
At the chosen conditions,
the system in the bulk exhibits two-phase coexistence for $x_A$ lying between $0.03$ and $0.97$.
For the confined system, we find $x_A^{\rm coex}=0.035$ [see Fig. \ref{fig7} (a)].

Fig.~\ref{fig7} shows examples for both $f (x_A, T)/k_BT$ [see part (a)]
and $\delta \mu (x_A, T)/k_BT$ [see part (b)]. The latter can be obtained from the
derivative of $f$ with respect to $x_A$. While in part (a)
we include results from three choices of $\varepsilon_a$,
including $\varepsilon_a=0$, part (b) contains results
only for $\varepsilon_a=0.1$, since chemical potential difference, vs. $x_A$, is
shown only to demonstrate the method. For both the quantities we have included
results only for a part of the overall $x_A$ range, that is relevant for the analysis.
It is clear from Fig. \ref{fig7} (a) that, as expected, with the increase of $\varepsilon_a$ the
energy barrier decreases, due to enhancement of favorable wetting condition.
The regions between the
first two knees of the free energy curves correspond to sphere-cap structure of droplets
rich in $A$ particles, radius of which increases with the increase of $x_A$.
By moving towards right one will encounter transitions to cylinder-like and slab-like structures,
below the wetting transition.
These are not of interest in this work. In this figure we have outlined
the estimation of the concentration difference $\Delta x$, from which, via a relation analogous to
Eq.~(\ref{eq14}), the actual contact angle is extracted. Specifically, $\Delta x$ is related
to the (wall attached) droplet volume ($V_d$) as
\begin{equation}\label{neq1}
\Delta x=(1-2x_A^{\rm coex})\frac{V_d}{V_{\rm box}}.
\end{equation}
Thus, one can
straightforwardly calculate $f_{VT}$, thus, $\Theta$, as a function of $x_A$, by obtaining information on the
volume reduction compared to a droplet in the bulk with same radius. This, of course, requires the
knowledge of $R$. Estimation of the latter we discuss below, though already described in
the previous section. In fact, missing information in Sec. III, if any, can be obtained from
the details below.

\begin{figure} [ht]
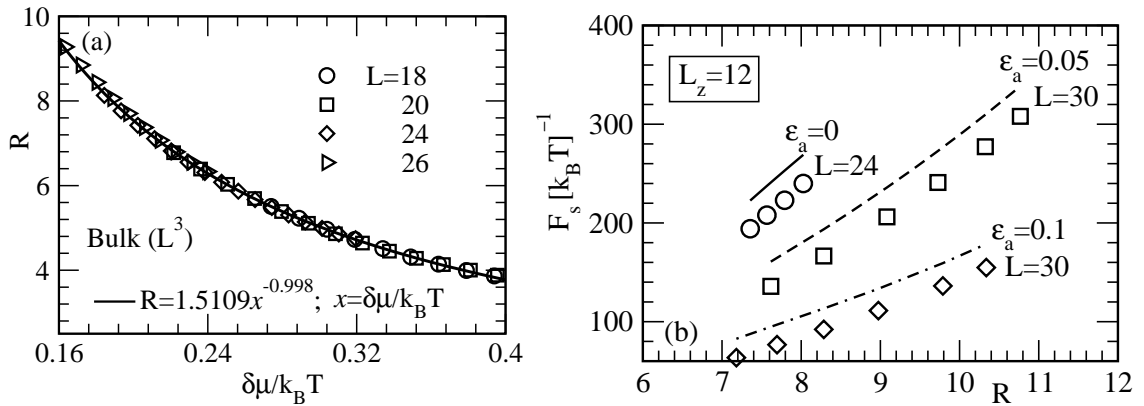

\includegraphics[width=6.9cm,clip]{fig8a.eps}
\includegraphics[width=8cm,clip]{fig8b.eps}
\centering
\caption{\label{fig8} (a) Variation of droplet radius $R$ with
chemical potential difference in the bulk symmetric binary Lennard-Jones fluid
(obtained by using cubic boxes of linear dimension
$L$ with periodic boundary conditions throughout; the chosen values of $L$ are
indicated in the figure). All data are compatible with the relation $R=1.5109$
$\Delta\mu^{-\eta}$ where the effective exponent is $\eta=0.998$. The radius is in
units of $\sigma$, and $\Delta \mu$ in units of $k_BT$. (b) Effective surface free
energy $F_s$ (in units of $k_BT$) is plotted versus the droplet radius $R$ 
for three choices of $\varepsilon_a$, as indicated. For $\varepsilon_a=0$, $L=24$ was used,
and $L=30$ for the other choices. The curves are the corresponding predictions
$F_s (R,\Theta_\infty)=F^{\rm sim}_s (R) f_{VT} (\Theta_\infty)$, with
$F^{\rm sim}_s (R)= 4 \pi R^2 \gamma_{AB}$.} \end{figure}

Due to the pronounced statistical fluctuations of the chemical potential difference,
the estimation of the latter, as well as of $\Delta x$, for
chosen states $x_A^{\rm box}$, has to be done with care. To facilitate working with smooth curves, we
have taken help of fitting of the simulation data to nonlinear functions.
E.g., the ascending branch of
$\delta \mu$ vs. $x_A$ plot was fitted to polynomial forms of degree four, while the descending
branch was fitted to power laws. For the corresponding radius
$R$ needed here, we use the relation between $R$ and $\delta \mu /k_BT$ found in the
bulk. Here we make use of the fact that droplets of same $R$ exist at same chemical
potential difference, irrespective of whether they are attached to the wall or not.
Related results are presented in Fig. \ref{fig8} (a). To avoid the finite-size
effects we have considered only overlapping data from different system sizes, as shown. For this purpose
also we have used the fitting method -- see caption for details.
There, essentially an inverse relation between $\delta\mu$ and $R$ emerges. This fact is consistent
with our discussion above [see Eqs. (\ref{eq7}) and (\ref{eq8}),
along with related text] with respect to vapor-liquid transition,
concerning the homogeneous nucleation theory for critical radius and corresponding energy barrier.
Even the prefactor ($\simeq 1.51$) is almost theoretically perfect, given that for
the binary mixture
\begin{equation}\label{neq2}
R=\frac{2\gamma_{AB}}{(1-2x_A^{\rm coex}) \delta\mu},
\end{equation}
where $\gamma_{AB}$ is $A-B$ interfacial tension.

Here we mention that in an earlier work \cite{20} we had obtained $R$ by using $\Theta (R=\infty)$,
calculated from an independent study that used a thermodynamic integration method, in the expression
$V_d=(4/3)\pi R^3 f_{VT} (\Theta)$. This and subsequent method did not naturally allow us to obtain any
curvature dependence of $\Theta$ and $\tau$.

From Fig.~\ref{fig7} (a) we can extract the actual (surface) excess free energy due to the
droplet, from the difference ($\Delta f$) between $f (x_A, T)/k_BT$ for the states at $x_A^{\nu}$ and
$x_A^{\rm box}$, as a function of $x_A^{\rm box}$, and thus, of $R$, given that $F_s=\Delta f V_{\rm box}$.
This can be compared with the result one would obtain if the line tension
effects were negligible [see lines of various types in Fig. \ref{fig8} (b)],
i.e., with $F_s/k_BT= 4 \pi R^2 \gamma_{AB} \, f_{VT} (\Theta)$,
using the actual contact angles found from the knowledge of $\Delta x$. The simulation results
for $F_s$ are presented as symbols
in Fig. \ref{fig8} (b). One sees that the actual surface
excess free energies are clearly smaller, indicating the presence of a negative line
tension.

\begin{figure} [ht]
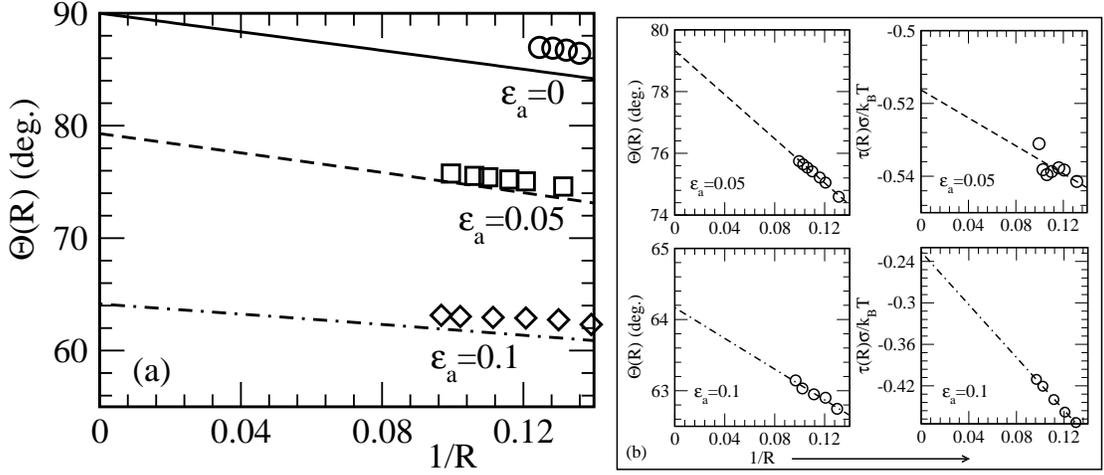

\includegraphics[width=8cm,clip]{fig9a.eps}
\includegraphics[width=6.5cm,clip]{fig9b.eps}
\centering
\caption{\label{fig9} (a) The lines are plots of the contact angle
$\Theta$, versus $1/R$, resulting from Eq.~(\ref{eq3}). For the case
$\varepsilon_a=0$, the line tension estimate (for planar interface) 
\cite{20} $\tau \sigma/k_BT=-0.52$ is used,
while in the other two cases the extrapolated results from part (b) of the figure are
used to draw the shown straight lines. Data shown by symbols are the values obtained from
the analysis of the MC simulation results.
(b) Simulation estimates for contact angles
$\Theta (R)$ (left) and line tension $\tau(R)$ (right) are plotted vs. $1/R$, for
$\varepsilon_a=0.05$ (upper frames) and $\varepsilon_a=0.1$ (lower frames). Straight lines
estimate $\tau$ and $\Theta$ at $R=\infty$ from (tentative) possible linear fits.
All results are from symmetric binary Lennard-Jones mixture.}\end{figure}

The line tension can be calculated from
the discrepancies between the continuous lines and the symbols in Fig. \ref{fig8} (b),
by using Eq. (\ref{eq10}).
Our best estimates for both contact angle and line tension are shown in
Fig.~\ref{fig9}, vs. $1/R$. In Fig. \ref{fig9} (a) we show contact angles from
three choices of $\varepsilon_a$. Note that for $\varepsilon_a=0$ one expects
$\Theta (R=\infty)=90^{\circ}$. In this figure we have also included the results that one
would obtain from Eq.~(\ref{eq3}). While Eq.~(\ref{eq3}) gives qualitatively
the right trend, a good quantitative agreement is not obtained. Unfortunately, only a rather
small range of values for $1/R$ is accessible for our analysis. Hence, as in the case
of the Ising model, rather tentative conclusions on the dependence of the line
tension on contact angle and radius can be inferred. However, we can conclude rather
safely that the contact angle for small wall-attached droplet is always significantly
smaller than its asymptotic value. Hence, as mentioned above, the results for the line tension obtained
previously in Ref. \cite{20}, for this model, suffer from systematic errors due to this
change of contact angle for small droplets.

We add here that for this model we have an estimate
of $\tau$ from planar interface available only \cite{20,49} for $\varepsilon_a=0$. For other values of
$\varepsilon_a$ we have obtained thermodynamic limit value of $\tau$ from extrapolations of
$R$-dependent data of $\tau$ to $R=\infty$. (Note that these plots are slightly
different from the Ising lattice gas case for which we have plotted $\tau$ as a function
of $1/r$, not $1/R$.) This procedure is shown in
Fig. \ref{fig9} (b) -- see the right frames. In the left frames of this figure we have shown
magnified pictures for the radius dependence of contact angle for the nonzero values of $\varepsilon_a$.
The values of $\Theta (R=\infty)$ obtained from here are used for the theoretical lines of Fig. \ref{fig9} (a).

\begin{figure} [ht]
\includegraphics[width=8cm,clip]{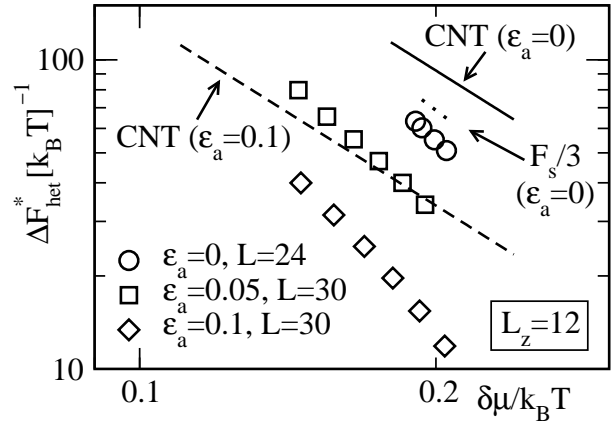}
\centering
\caption{\label{fig10} Plots of nucleation barrier as a function of chemical potential,
for the symmetric binary Lennard-Jones fluid confined between antisymmetric walls. 
Simulation results (symbols) for a few
different values of $\varepsilon_a$ are included. For $\varepsilon_a=0$ and $0.1$ we have
shown the predictions of classical theory (marked as CNT) as well. For $\varepsilon_a=0$ we have
included a plot for $F_s/3$.} \end{figure}

Before closing this section, in Fig. \ref{fig10} we present results for the barriers for
heterogeneous nucleation, i.e., we plot $\Delta F_{\rm het}^*/k_BT$ versus $\delta\mu/k_BT$,
for different values of $\varepsilon_a$. These results
are analogous to Fig. \ref{fig6} (that corresponds to 
the lattice gas model). Here also we compare the simulation
results with the classical theory (marked as CNT). The discrepancies between theory and simulation
again imply negative line tension. Here we have not included the entropic contribution to $F_s$.

\section{Mean-field type calculation for the dependence of contact angle on the chemical
potential difference}

In this section, we return to the Ising model between antisymmetric walls,
and reinterpret the two spin values as particles of type A and B. Being interested
in temperatures far below the bulk critical temperature, it is tempting to study
contact angles within a mean-field type approximation. This is to obtain
clarification to what extent the variation of the line tension with droplet radius,
implied by the results of the previous sections, is a fluctuation effect, or
simply an effect due to the dependence of wall free energies on the chemical potential
difference between the droplet and its environment.

A convenient construction of mean-field theory for inhomogeneous lattice problems
is the Scheutjens-Fleer formulation \cite{46} of self-consistent field theory (SCFT).
While SCFT was originally intended for polymeric systems, it can be applied to a
lattice model for binary mixtures as well. Since SCFT directly yields the
free energy of the system, one can obtain wall excess free energies straightforwardly
\cite{26,46}. Denoting the normalized interaction energy (Flory-Huggins parameter) as
\begin{equation} \label{eq26}
\chi_{AB} = q \frac{[\varepsilon_{AB} - 0.5(\varepsilon_{AA} + \varepsilon_{BB})]}{k_BT},
\end{equation}
where $q=6$ is the coordination number of the simple cubic lattice, we note that
criticality in the bulk would occur for $\chi^{\rm crit}_{AB}=2$. So the regime of
interest is $\chi_{AB} > 2$.
Similarly, one can define the normalized interaction energy of $A$-particles with $S$-particles forming the wall as 

\begin{equation} \label{eq26surface}
\chi_{AS} = q \frac{[\varepsilon_{AS} - 0.5(\varepsilon_{AA} + \varepsilon_{SS})]}{k_BT},
\end{equation}
and likewise, the normalized interaction energy of $B$-particles with $S$-particles ($\chi_{BS}$) can be defined by replacing $A$ with $B$ everywhere in the above equation.

\begin{figure} [ht]
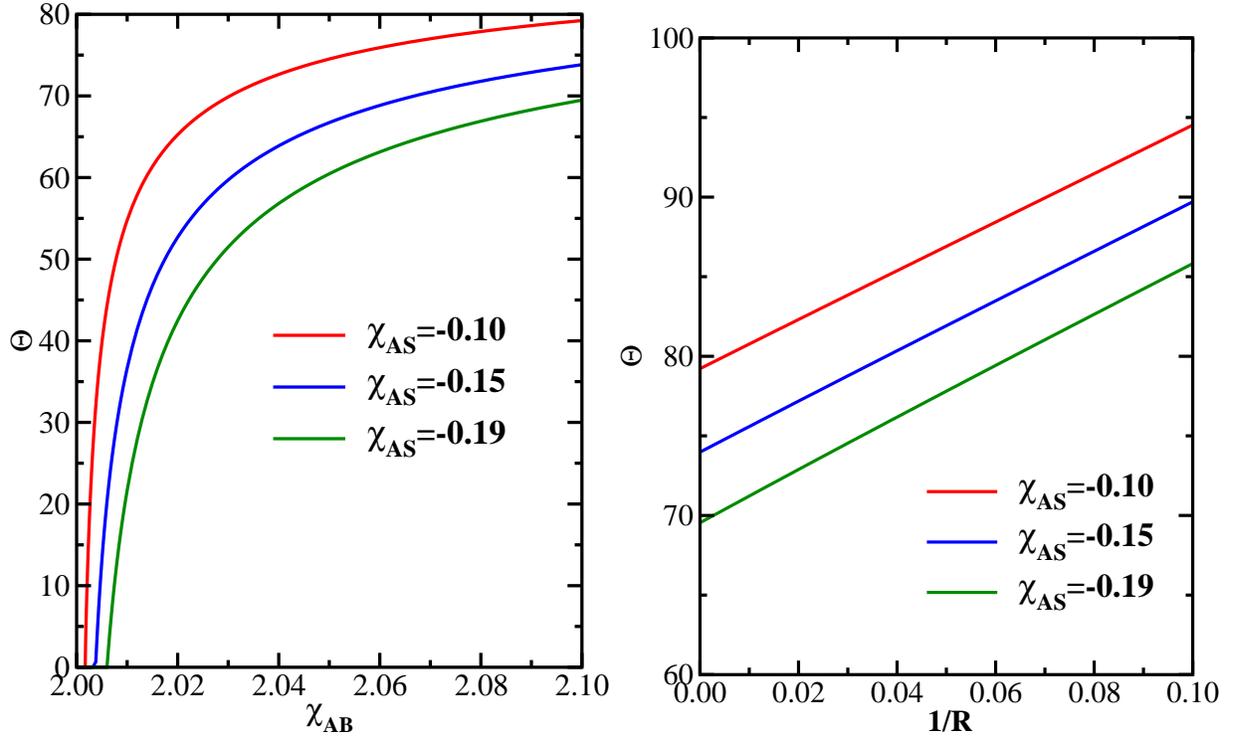

\includegraphics[width=8cm,clip]{fig11a.eps}
\includegraphics[width=8cm,clip]{fig11b.eps}
\centering
\caption{\label{fig11} a) Contact angle as a function of the (Flory Huggins) parameter
$\chi_{AB}$ for the mean-field model, for three values of the wall attraction
strength $\chi_{AS}$. b) Contact angle, computed
from Eq.~(\ref{eq29}), is plotted vs. $1/R$, for $\chi_{AB}=2.1$
and the same three values of $\chi_{AS}$ as shown in part (a).
} \end{figure}

Using walls, which attract either $A$-particles with
attraction strength $\chi_{AS}$ or $B$-particles with attraction strength $\chi_{BS}$,
one can both obtain wall excess free energies at coexistence conditions, and study
also antisymmetric systems with an inclined interface, similar to M\"uller et al.
\cite{26}. Using Young's equation, Eq.~(\ref{eq1}), one readily obtains the contact
angle $\Theta_\infty$ as a function of these energy parameters (see Fig.~\ref{fig11} (a) as an example).
One sees that $\Theta_\infty$ vanishes when the wetting transition occurs but for $\chi_{AB}=2.1$ already rather
large values of $\Theta_\infty$ occur. Hence we shall focus on the $R$-dependence of $\Theta$ for this case in the
following.

For the calculation of wall tensions $\gamma_{wv}(\delta \mu)$ and
$\gamma_{w \ell} (\delta \mu)$ from this mean-field theory, one uses a
$L \times L \times L_z$ geometry, with periodic boundary conditions invoked in
$x$ and $y$ directions, and on a mean-field level only inhomogeneity
in $z$-direction occurs. Thus the implementation of SCFT for this calculation
is rather straightforward.

In order to carry out a meaningful comparison with the result of the previous
sections, we assume chemical potential differences related to a droplet radius
according to Eq. (\ref{neq2}). For the chosen example (Fig.~\ref{fig11}) with $\chi_{AB} =2.1$ we have
$\gamma_{AB}=0.0123$  and $1-2 x_A^{\rm coex}=0.37$. Being only
interested in the leading behavior for large $R$, it suffices to expand the
wall tensions at the coexistence curve according to Eqs. (\ref{nneqn4}) and (\ref{nneqn5})
by identifying the liquid phase with $A$-rich phase and the vapor phase with the $B$-rich phase.
From Eqs.~(\ref{neq2}), (\ref{nneqn4}) and (\ref{nneqn5}) we immediately conclude
\begin{equation} \label{eq29}
\cos \Theta=\frac{\gamma_{wB}-\gamma_{wA}}{\gamma_{AB}}= \cos \Theta_\infty + \frac{2 \Delta \Gamma}{(1-2x_A^{\rm coex})}
\, \frac{1}{R},
\end{equation}
where $\Delta\Gamma= \Gamma_A-\Gamma_B$, computed at coexistence. Note that Eq.~(\ref{eq29}) holds beyond 
mean field, for $R \rightarrow \infty$, if the dependence on wall excess free energies on $\delta\mu$ is the 
only reason for a difference between $\Theta$ and $\Theta_\infty$.

In Eq. (\ref{eq29}), $\cos \Theta_\infty$ simply is the result according to Young's equation,
applying for $R \rightarrow \infty$. Clearly, mean-field theory predicts that
the dependence of the contact angle on the chemical potential difference
$\delta \mu$, and hence via Eq.~(\ref{neq2}) indirectly on $R$, is a rather
pronounced effect! This finding confirms the concerns raised by Schimmele
and Dietrich \cite{11} that one has to be very careful when one wishes to
associate the radius dependence of the contact angle of wall attached sphere-cap
shaped droplets with the ``true'' line tension.

It is tempting to use a higher-dimensional version of this mean-field theory
for computing the line tension \cite{47}. However, due to lattice effects
(interfaces in mean-field theory stay non-rough up to bulk criticality) this
is rather difficult to implement, and hence not done here.

\section{Discussion}
In this paper, we have presented a re-analysis of simulation data, that
were used in two earlier attempts
\cite{18,19,20}, to extract estimates for the contact angle $\Theta$ of
wall-attached sphere-cap shaped droplets and
the line tension related to the corresponding three-phase contact,
by paying attention to a dependence on the
droplet radius. In the previous works, a heuristic assumption was made
that the contact angle of these droplets is the same as the contact angle
$\Theta_\infty$ of macroscopic droplets, which can be estimated independently
from Young's equation. As a matter of fact there is no theoretical basis for
this assumption, and hence it is clearly desirable to avoid it. We show
here that this task can indeed be achieved by the simulation strategy of the previous
work \cite{18,19,20}, by using the explicit knowledge of the relation between
the droplet radius of curvature $R$ and the chemical potential difference
$\delta \mu$ that characterizes the equilibrium between the system containing
the droplet and bulk phase coexistence, and the excess density due to the
droplet.

However, a crucial approximation which remains is that for the excess volume of the
wall-attached droplet a sphere-cap shape is accurate, i.e., corrections to
Eq.~(\ref{eq4}) due to deviations of the droplet shape near the contact line
are sufficiently small so that they can be neglected. This is plausible since only a volume
of order $2 \pi R a^2 \sin \Theta$, where $a$ is a length of the order of molecular
distances, should be affected. Hence for the volume of the sphere cap this
is a correction of order $(a/R)^2$. So we expect a correction for our
contact angle of order $(a/R)^2$, while the corrections that are
found and discussed here are of order $1/R$, much stronger than above errors can bring in.
In the case of the lattice gas model, a further caveat is that the anisotropy
of the lattice also causes systematic deviations of droplet shapes from being
spherical (or sphere-cap, respectively). But at the chosen temperature these
anisotropy effects should be very small.

If these assumptions are accepted, direct estimation of $\Theta$ as a function
of $R$ is possible. We have interpreted the deviations between $\Theta$
and $\Theta_\infty$ in terms of line tension effects,
notwithstanding the knowledge that in view of the criticisms
raised by Schimmele et al. \cite{10,11} this is doubtful. For the Ising model,
an estimation of the dependence of the ``true'' line tension (referring to
macroscopic planar interfaces) on contact angle is available \cite{23}, and
this dependence is incompatible with previous estimates resulting from droplets
\cite{18,19}. This led to a motivation for the present work. However, the present
results are hardly compatible with the implicit assumption of \cite{15}, viz.,
the line tension is a true constant, i.e., $\tau$ does not depend upon $R$
and $\Theta$. We have, however, provided evidence, already
on the mean-field level, that the dependence of contact angles
(even for $R \rightarrow \infty$) on the chemical potential $\delta\mu$ in
the system (for finite $R$, $\delta\mu$ must be nonzero in equilibrium) is of
a similar magnitude as the dependence attributed to line tension effects.

The above findings agree with concerns raised by Schimmele and Dietrich \cite{10,11}.
Of course, computer simulations suffer from well-known problems such as finite
size effects, statistical errors, incomplete equilibration, etc. \cite{42}.
Hence it is not straightforward to clarify all the issues that our work raises.
However, other numerical approaches such as density functional theory \cite{14}
seem to provide a rather strong variation ($\tau \propto R)$ for small $R$,
and also raise questions.

In conclusion, although the concept of the line tension
was introduced by Gibbs more than a century ago \cite{7}, its numerical estimation
by experiment, simulation and theory still is difficult. An interesting point is
that for both the models, lattice gas and binary Lennard-Jones,
the variation of the contact angle of droplets seems
to be compatible with a linear behavior in $1/R$.

Finally, we again like to emphasize that our simulation approach yields directly estimates for
the barrier $\Delta F^*_{\rm het}$ that needs to be overcome in the heterogeneous nucleation events,
which are not at all affected by uncertainties in our knowledge of contact angles and line
tensions. These estimates (e.g. Fig.~\ref{fig6} and Fig. \ref{fig10}) 
do not make any assumption what the actual
droplet shapes are, but rely fully on our estimates of the excess density due to the droplet and
associated free energy excess.

\section*{Acknowledgment} SKD acknowledges the Marie Curie Actions Plan of European 
Commission (FP7-PEOPLE-2013-IRSES grant No. 612707, DIONICOS), International Centre for 
Theoretical Physics, Trieste, and Johannes-Gutenberg University of Mainz for partial supports.

\clearpage

\end{document}